\documentclass[onecolumn,superscriptaddress,secnumarabic,amssymb, nobibnotes, aps, prf, 12pt]{revtex4-2}

\setcitestyle{numbers,square}
\usepackage[colorlinks=true,linkcolor=green, citecolor=green,allcolors=blue]{hyperref}

\usepackage{graphicx}
\usepackage{dcolumn}
\usepackage{bm}
\usepackage[utf8]{inputenc}
\usepackage[T1]{fontenc}
\usepackage{mathptmx}
\usepackage{etoolbox}
\UseRawInputEncoding

\makeatother
\usepackage{graphicx}
\usepackage{amsmath}
\usepackage{mathtools}
\usepackage{upgreek}
\usepackage{hhline}
\usepackage{here}
\usepackage{booktabs}
\usepackage{multirow}
\usepackage{array}   
\usepackage{tabularx}
\usepackage{textcomp,gensymb}
\usepackage[per-mode=symbol,range-phrase=--,range-units=single]{siunitx}
\usepackage{color}
\usepackage{nicefrac}
\usepackage{xcolor}
\usepackage{cleveref}
\usepackage[]{changes}
\usepackage{setspace}
\hypersetup{
    colorlinks=true,
    linkcolor=blue,
    filecolor=blue,      
    urlcolor=blue,
}

\newcommand{\ARCNL}{Advanced Research Center for Nanolithography (ARCNL),\\Science Park 106, 1098 XG Amsterdam, The Netherlands}
\newcommand{\VU}{LaserLab, Department of Physics and Astronomy, Vrije Universiteit Amsterdam,\\ De Boelelaan 1100, 1081 HZ Amsterdam, The Netherlands}
\newcommand{\UvA}{Van der Waals\textendash Zeeman Institute, Institute of Physics, University of Amsterdam, \\ 1098XH Amsterdam, The Netherlands}

\newcommand{\We}{\mathrm{We}}
\newcommand{\Wed}{\mathrm{We_d}}
\newcommand{\Weu}{\mathrm{We}_\mathrm{\hat{U}}}
\newcommand{\W}{\mathrm{W}}
\newcommand{\tauc}{\tau_\mathrm{c}}

\begin{document}
\singlespacing 
\title{Singular jets in free-falling droplets}

\author{M. Kharbedia}
\email{kharbedia@arcnl.nl}
\affiliation{\ARCNL}

\author{H. Franca}
\affiliation{\ARCNL}
\affiliation{\UvA}

\author{H. K. Schubert}
\affiliation{\ARCNL}
\affiliation{\VU}

\author{\mbox{D. J. Engels}}
\affiliation{\ARCNL}
\affiliation{\VU}

\author{M. Jalaal}
 \affiliation{\UvA}

\author{O. O. Versolato}
\email{versolato@arcnl.nl}
\affiliation{\ARCNL}
\affiliation{\VU}

\date{\today}

\begin{abstract}
We report on singular jets in a free-falling liquid tin droplet following nanosecond laser-pulse impact. 
Following impact, the droplet (with diameter $D_0=50$ or 70\,$\mu$m) undergoes rapid radial expansion and subsequent retraction, resulting in the formation of an axisymmetric jet. 
Using numerical simulations in tandem with our experiments, we reveal that a delicate interplay between radial flow and the curvature of the retracting droplet governs jet formation. 
The resulting dynamics is characterized using the impact Weber number, $\We$ (in the experiments $2 \lesssim \We \lesssim 16$), and a pressure width, W (typically $1 \lesssim \W \lesssim 2$), which describes the angular distribution over the droplet surface of the instantaneous pressure impulse exerted by the transient laser-produced plasma. 
For values $\We<10$, the droplet presents a pronounced forward curvature during the retraction, leading to the formation of a cavity. 
The collapse of such a cavity leads to a singular jet that greatly enhances the jetting velocity up to ten times the impact propulsion velocity, an effect that narrowly peaks around $\We\sim6-8$, reminiscent of singular jets in droplet-solid impact. 
We identify a further sensitivity of the jet velocity enhancement on the pressure width W and capture the dynamics in a phase diagram connecting the various deformation morphologies with jet velocity.

\end{abstract}\def\keywordstitle{Keywords}

\maketitle

\onecolumngrid

\section{Introduction}\label{Introduction}

Liquid jet formation at the macroscale is observed across a wide range of flow conditions. Notable examples include droplet impact on a solid surface\,\cite{bartolo_singular_2006,ding2012propagation,josserand_drop_2016,chen_submillimeter-sized_2017,zhao_jetting_2020,guo_oblique_2020,zhang2022impact,sanjay_role_2025} and on a liquid bath\,\cite{Worthington_drawing,castillo2015droplet,thoroddsen2018singular_jet,blancorodriguez_jets_2021, tian_throddsen2023conical_singular_jet}, bubble bursting at a liquid interface\,\cite{duchemin2002jet,lee_size_jet_2011, ghabache2016size, deike2018dynamics, Sanjay2021, gordillo2023theory_jets, balasubramanian2024bursting}, and cavity collapse resulting from the impact of a solid object on a liquid surface\,\cite{Worthington_drawing,gekle2009high, gekle_gordillo_2010, gordillo2019capillary}. These jetting phenomena have important industrial applications, such as inkjet printing\,\cite{Lohse_inkjet_2022, Liu2023_Inkjet}, laser-induced forward transfer~\cite{serra2019laser,jalaal2019laser,jalaal2019destructive}, and needle-free medical injection devices\,\cite{kale2014needle, barolet2018current_needle_free}. They also embody complex fluid dynamics present in nature, for instance, in raindrop impacts on leaves\,\cite{lejeune2019drop_plant}, or aerosol exchange between ocean and atmosphere during wave breaking\,\cite{veron2015ocean_spray,berny2021statistics}.

The jetting dynamics observed in droplet impact on non-wetting surfaces is of particular interest in this study. At sufficiently low impact velocities, the radially converging surface waves upon impact lead to the formation of an air cavity that collapses during the droplet's recoil phase\,\cite{bartolo_singular_2006, thoroddsen_microjetting_2007, thoroddsen2009, chen_submillimeter-sized_2017, thoroddsen2018singular_jet, zhang2022impact, sanjay_role_2025, Singular_jet_compount_drop_2026}. For low radial flows the droplet retracts and breaks axially, while for higher radial flows the initial spreading leads to the formation of a thin film. 
Within a narrow range of impact velocities, the cavity collapse and droplet recoil can lead to a fast singular jet, often with a velocity several times greater than the initial impact velocity. Similar jetting behavior is observed when solid objects fall into liquid pools\,\cite{gekle_gordillo_2010, blancorodriguez_jets_2021}. 
The emergence of such jets critically depends on the convergence of surface capillary waves generated after the impact, resulting in a toroidal droplet shape, 
combined with the radial recoil motion, both of which enhance the jetting process\,\cite{bartolo_singular_2006, chen_submillimeter-sized_2017, yamamoto_initiation_2018}. On non-wetting hydrophobic or superhydrophobic surfaces, bubbles may become trapped beneath the droplet during impact. These bubbles can migrate upward and burst at the surface, also producing high-speed jets\,\cite{bartolo_singular_2006, chen_submillimeter-sized_2017,peng_2025}. Wettability further influences the dynamics: On certain surfaces, the entrapped cavity may become pinned, resulting in asymmetric closure. This asymmetry can promote necking and pinch-off of the cavity, significantly amplifying jet velocities\,\cite{bartolo_singular_2006, chen_submillimeter-sized_2017}. Furthermore, highly focused microjets produced within a closing cavity after droplet impact on a liquid pool can interact with the surrounding air, influencing the jet velocity due in the confined geometry\,\cite{tian_throddsen2023conical_singular_jet}. Together, these phenomena illustrate a complex interplay between the radial flow within the droplet, the cavity closure mechanism, and the propagation and focusing of capillary waves, leading to jet formation.

In this work, we find singular jets arising from the interaction of a nanosecond laser pulse with a free-falling liquid tin droplet in vacuum. Here, the droplet undergoes rapid expansion due to the recoil pressure of the plasma formed on the laser-facing surface\,\cite{kurilovich2018power, hernandez2022early}. 
The relevant parameters used to characterize these dynamics include the Weber number based on the center-of-mass propulsion velocity, $U_\mathrm{cm}$, defined as $\We=\rho D_0U_\mathrm{cm}^2/\sigma$; deformation Weber number based on the radial expansion rate $\dot{R}$, defined as $\Wed=\rho D_0\dot{R}/\sigma$ following Ref.\,\cite{klein_drop_2020}; and a pressure distribution width $\W$, modeled as a raised cosine function $\propto \left[1-\cos\left(\theta\frac{\pi}{\W}\right)\right]$, with azimuthal angle $\theta$, to represent the angular distribution of the applied surface pressure following Ref.\,\cite{francca2025laser}. 
Here, $D_0$ is the initial droplet diameter, $\rho=7000\,\mathrm{kg/m^3}$ is the density of liquid tin, and $\sigma=0.544\,\mathrm{N/m}$ is the surface tension\,\cite{Sn_viscosity}. For low $\We$, the droplet undergoes radial expansion followed by contraction, resembling the recoil-driven flow observed in droplet impact scenarios. Due to the angular distribution of the surface pressure, the droplet acquires an effective curvature during retraction\,\cite{hernandez2022early, francca2025laser}. This curvature can facilitate the entrapment of a cavity (containing a vacuum), which, upon collapse, produces a strong jet directed along the laser propagation axis. Additionally, based on the jet's tip velocity, $\hat{U}$ we define the corresponding jetting Weber number $\Weu=\rho D_0 \hat{U}^2/\sigma$ that permits us to define breakup criteria.   
The absence of a substrate and a contact line offers an ideal system for droplet flow analysis, eliminating any dependence of the jetting dynamics on substrate wettability. Moreover, the drag forces that arise from interactions with the surrounding air are minimal as experiments are performed in vacuum. These conditions enable the production of singular jetting within a remarkably simplified system, where an effectively instantaneous touch of a laser fully determines the droplet dynamics. The additional control parameter (that is, besides $\We$) in this laser-droplet interaction is the width of the spatial distribution of the recoil pressure generated by the plasma formed at the droplet surface, denoted by $\W$\,\cite{kharbedia2025_LWO, francca2025laser}.

The two governing parameters, $\We$ and $\W$, can be tuned via the laser energy\,\cite{francca2025laser,kharbedia2025_LWO}, leading to a precise control of the radial flow and the curvature of the retracting droplet. For a fixed beam size as in the case of our experiments (see Ref.\,\cite{hernandez2022early} for the influence of the size of the beam focus), these two parameters are directly linked, leading to a broader pressure field by increasing propulsion Weber number\,\cite{kharbedia2025_LWO} through increasing the laser pulse energy. 
At low Weber numbers ($\We < 6$), the radial expansion is limited ($D_\mathrm{max} \sim D_0$), and the droplet primarily deforms along the laser propagation axis, conferring onto the retracting droplet a pronounced curvature that favors the entrapment of a cavity. As $\We$ increases, the radial flow becomes more pronounced, creating favorable conditions for singular cavity collapse\,\cite{bartolo_singular_2006}, particularly around $\We \sim 6-8$. While the radial flow increases with $\We$, the retraction becomes more forward-backward symmetric because the pressure distribution broadens ($\W$ increases) for higher $\We$ values. At higher Weber numbers $\We>10$, cavity formation is suppressed. Consequently, the observed jet becomes thicker and slower. Therefore, we identify a narrow range of Weber numbers ($6<\We<8$) in which the combination of sufficient radial flow and pronounced curvature leads to the formation of a high-speed, singular jet. To further investigate the cavity dynamics underlying this jetting mechanism, we complement our experiments with numerical simulations. We observe complex cavity behavior, ranging from symmetric to asymmetric collapse, followed by microscopic bubble entrapment within the droplet. With these results, we pave the way for precise control of microjetting in free-falling droplets by tuning both the radial flow and the droplet curvature through the laser parameters.

In the following, we briefly present the experimental and numerical methodologies employed in this study (\S\ref{sec:methods}). We then present the most representative results related to the enhanced jetting mechanism (\S\ref{sec:res}), with a primary focus on the comparison between simulations and experiments. First, we analyze the jetting velocity and its dependence on the Weber number $\We$ (\S\ref{sec:cavity}). Next, using numerical simulations, we study separately the influence of the angular distribution of the recoil pressure on the jetting process, which allows us to build a phase diagram in terms of $\W$ and $\We$, where we show different cavity collapse mechanisms that determine jetting dynamics (\S\ref{sec:phase_diagram}). Finally, we conclude by presenting a particular case in which the retracting sheet develops radially converging surface waves, affecting the resulting jet velocity (\S\ref{sec:stepwise}).     

\begin{table*}[t]
    \centering
    \caption{Relevant experimental parameters: Droplet diameter $D_0$, capillary timescale $\tauc=\sqrt{\rho D_0^3/6\sigma}$, pulse energy $E_\mathrm{p}$, pressure width $\W$, center-of-mass propulsion rate $U_\mathrm{cm}$, propulsion Weber number $\We=\rho D_0 U_\mathrm{cm}^2/\sigma$, radial expansion rate $\dot{R}_0$, deformation Weber number $\Wed=\rho D_0 \dot{R}_0^2/\sigma$, jet velocity $\hat{U}$, and the corresponding Weber number $\Weu=\rho D_0 \hat{U}^2/\sigma$.}
    
    \renewcommand{\arraystretch}{1.1}
    \setlength{\tabcolsep}{8pt}
    
    {\fontsize{10}{14}\selectfont 
    
    \begin{tabular}{cccccccccc}
        \hline
        $D_0$ ($\mu\mathrm{m}$) & 
        $\tauc$ ($\mu\mathrm{s}$) & 
        $E_\mathrm{p}$ (mJ) & 
        $\W$ & 
        $U_\mathrm{cm}$ (m/s) & 
        $\We$ & 
        $\dot{R}_0$ (m/s) & 
        $\Wed$ & 
        $\hat{U}$ (m/s) & 
        $\Weu$ \\
        \hline
        50 & 16.4 & 0.5--2.4 & 1.4--1.8 & 2.2--7.4 & 3--14 & 4.1--11 & 10--43 & 4.3--13 & 12--194 \\
        70 & 27.1 & 0.6--4.0 & 1.3--1.9 & 1.5--8.7 & 2--16 & 2.8--9.5 & 3--41 & 3.8--15 & 5--590 \\
        \hline
    \end{tabular}
    
    }
    \label{tab:table1}
\end{table*}


\section{Method}\label{sec:methods}

\subsection{Experiment}\label{subsec:exp}

\begin{figure}[h!]
    \centering
    \includegraphics[width=1\textwidth]{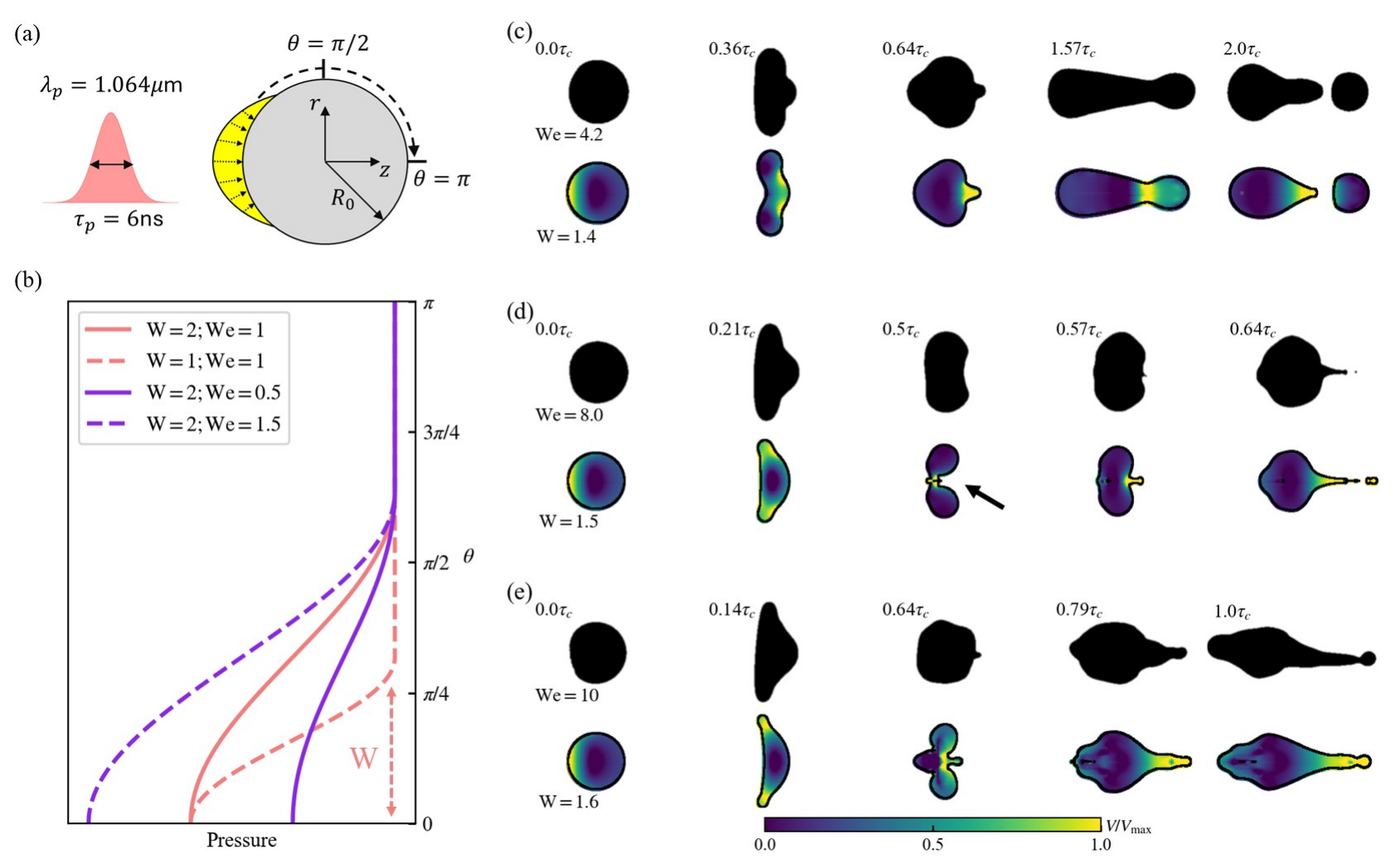}  
    \caption{Comparison of experiment and simulation. (a) Conceptual sketch depicting a Gaussian nano-second laser hitting the droplet with radius $R_0$ from the left. The resulting plasma after surface ablation is depicted with a yellow evolving area, where the dashed arrows indicate the plasma recoil pressure. (b) Examples of angular distributions of the plasma pressure on the droplet's surface for two different $\We$ values at fixed $\W$ (solid lines) and two different $\W$ values for fixed $\We$ (dashed lines). Conceptual illustration of the pressure width is indicated with the dashed arrow. (c-e) Comparison of the experimental data (black shapes) for droplet jetting with numerical simulations (colored shapes) performed at the same values of $\We$ over different fractions of capillary time $\tauc$; the values for $\W$ are estimated from the empirical scaling (see main text). The $\We$ increases from top to bottom. The $1^{\mathrm{st}}$ frame shows the droplet at rest; the $2^{\mathrm{nd}}$ frame shows the droplet at its maximum radial extension $\sim D_\mathrm{max}$; The $3^{\mathrm{rd}}$ frame depicts the jet emerging from the droplet after its contraction. The $4^{\mathrm{th}}-5^{\mathrm{th}}$ frames illustrate the jet evolution over time. The maximum jetting velocity is observed at $\We=8$. Note the presence of a cavity developed during retraction for $\We=8$ at $0.5\tauc$, which produces a singular jet (indicated with the arrow; see main text for discussion).}
    \label{fig:presentation}
\end{figure}

For a detailed description of our experimental setup, we refer to Refs.\,\cite{Kurilovich2018ps, Meijer2022_transition, liu_2023_mass, kharbedia2025rim}. 
Briefly, we produce liquid tin microdroplets with diameters $D_0$=50 or 70$\,\mu\mathrm{m}$ in a vacuum chamber with a base pressure $10^{-6}$\,mbar. 
A circularly polarized Nd:YAG laser pulse hits the droplet with pulse energies within the range $0.5-4\,\mathrm{mJ}$ (see Table\,\ref{tab:table1}). The laser is focused on the droplet as a Gaussian spot with $\sim85\,\mu\mathrm{m}$ diameter at full width at half maximum (FWHM). 
We use a stroboscopic imaging system to visualize the laser-droplet interaction, recording the shadow from a $90\degree$ angle (that is, orthogonal to the propagation direction of the laser beam), after illuminating the droplet with a temporally and spatially incoherent green light.
The optical resolution achieved with this setup is approximately $\sim5\,\mu\mathrm{m}$.
The images are captured with a CCD camera. 
A delay generator is used to precisely control the timing of the laser impact on the droplet and its subsequent illumination for imaging.
To observe the time evolution of the droplet, the shadowgraph illumination pulse is delayed with respect to the laser pulse from 0 to $\sim$50\,$\mu$s, typically in 1.1\,$\mu$s steps.
     
\subsection{Simulation}\label{sec:sim}

We perform numerical simulations to study the time evolution of the droplets after laser impact. We make use of the Navier-Stokes equations for an incompressible, isothermal bi-phase (droplet and surrounding low density ambient) flow, where the momentum and continuity equations inside the droplet phase are:

\begin{gather}
    \frac{\partial {\textbf{u}}}{\partial t} +  ({\textbf{u}} \cdot \nabla) {\textbf{u}}= \textbf{f}_\sigma - \frac{1}{\rho} \nabla p + \frac{\mu}{\rho} \nabla^2\textbf{u}, \label{eq:NS-1} \\
    \nabla \cdot \,\textbf{u} = 0, \label{eq:NS-2}
\end{gather}
where \textbf{u} and $p$ are the velocity and pressure fields, $\mu$ is the dynamic viscosity, and $\textbf{f}_\sigma=\sigma\kappa\delta_\mathrm{k}\textbf{n}$ performs the role of the surface tension as a body force with $\kappa$ being the local curvature on the interface between the droplet and the ambient fluid, $\delta_\mathrm{k}$ is the Dirac delta function, and \textbf{n} is the unit vector normal to the interface\,\cite{popinet2009accurate, Tryggvason2011-book}. 
The following nondimensionalization is performed of the governing equations: 

\begin{align}
    \textbf{x} &= R_0\bar{\textbf{x}}, & 
    t &= \frac{R_0}{U}\bar{t}, & 
    \textbf{u} &= U\bar{\textbf{u}}, \nonumber \\
    p &= \rho U^2\bar{p}, & 
    \kappa &= \frac{1}{R_0}\bar{\kappa}, & 
    \delta_s &= \frac{1}{R_0}\bar{\delta_s}\,.
    \label{eq:nondim_scales}
\end{align}
giving rise to the dimensionless version of Eqs.\,(\ref{eq:NS-1},\ref{eq:NS-2}) inside the droplet phase:

\begin{gather}
    \frac{\partial \bar{\textbf{u}}}{\partial t} +  (\bar{\textbf{u}} \cdot \nabla) \bar{\textbf{u}} = - \nabla \bar{p}  + \frac{1}{\textrm{Re}} \nabla^2\bar{\textbf{u}}  + \frac{1}{\textrm{We}} \bar{\kappa} \bar{\delta_s} \textbf{n}, 
    \label{eq:NS_nond1} \\
    \nabla \cdot \bar{{\textbf{u}}} = 0, \label{eq:NS_nond2}
\end{gather}
where $\mathrm{Re}=\rho D_0 U/\mu$ is the Reynolds number. Note that in the experiment only $U$ is changed, resulting in the changing of both Re and We, hence in simulations we obtain the corresponding Re number for each We (to range $\sim$100--1000). 
We solve Eqs.\,(\ref{eq:NS-1}-\ref{eq:NS_nond2}) numerically with the open-source code Basilisk C\,\cite{Popinet2013-Basilisk}. Based on mesh refinement, we resolve the flow described by Eqs.\,(\ref{eq:NS_nond1}-\ref{eq:NS_nond2}), with the droplet at the center of a box with dimensions [$-5D_0$,$5D_0$] on both $r$ and $z$ axis [see Fig.\,\ref{fig:presentation}(a)]. Furthermore, the interface between the droplet and the ambient fluid is represented following the VOF scheme\,\cite{Hirt1981}, in which the fraction of the droplet fluid is indicated by a scalar function for each cell in the simulation. In this bi-phase model, we define the surrounding fluid density and viscosity as $\rho_\mathrm{a}=10^{-4}\rho_\mathrm{d}$, and $\mu_\mathrm{a}=10^{-4}\mu_\mathrm{d}$, respectively, where the indices a,d refer to ambient and droplet with $\mu=1.8\times10^{-3}\,\mathrm{Pa.s}$ being the liquid tin viscosity\,\cite{Density_tin_Asseal}. A set of equations similar to Eqs. \eqref{eq:NS_nond1}-\eqref{eq:NS_nond2} is solved on the ambient phase, but using these lower density $\rho_\mathrm{a}$  and viscosity $\mu_\mathrm{a}$. Despite the fact that, under our experimental conditions, the density ratio is much smaller, $\rho_\mathrm{a}=10^{-13}\rho_\mathrm{d}$, the numerical value used remains sufficiently low to adequately reproduce the experimental observations (see Franca et al.\,\cite{francca2025laser}). Although the liquid viscosity is kept constant in all simulations, further details on its influence on the jetting dynamics are provided in Appendix\,\S\,\ref{app:viscosity}. We parametrize the pressure profile imprinted by the laser on the droplet's surface with a raised cosine $P=\frac{1}{2}\left[1-\cos\left(\theta\frac{\pi}{\W}\right)\right]\,H\left(\W-\theta\right)$, with $\theta$ as the azimuthal angle, as illustrated in Figs.\,\ref{fig:presentation}(a,b), and $H$ the usual Heaviside function. 
Here, $\W$ is again the width of the pressure field on the droplet. 
Fig.\,\ref{fig:presentation}(b) presents example pressure distributions. 
Note that in experiments the weber number $\We$ and pressure width $\W$ are correlated as described in more detail in Ref.\,\cite{kharbedia2025_LWO}. 
A more detailed explanation of the numerical methodology is provided in Ref.\,\cite{francca2025laser}.

\section{Results and discussion}\label{sec:res}
\subsection{Cavity entrapment and droplet jetting}\label{sec:cavity}

\begin{figure}
    \centering
    \includegraphics[width=0.9\textwidth]{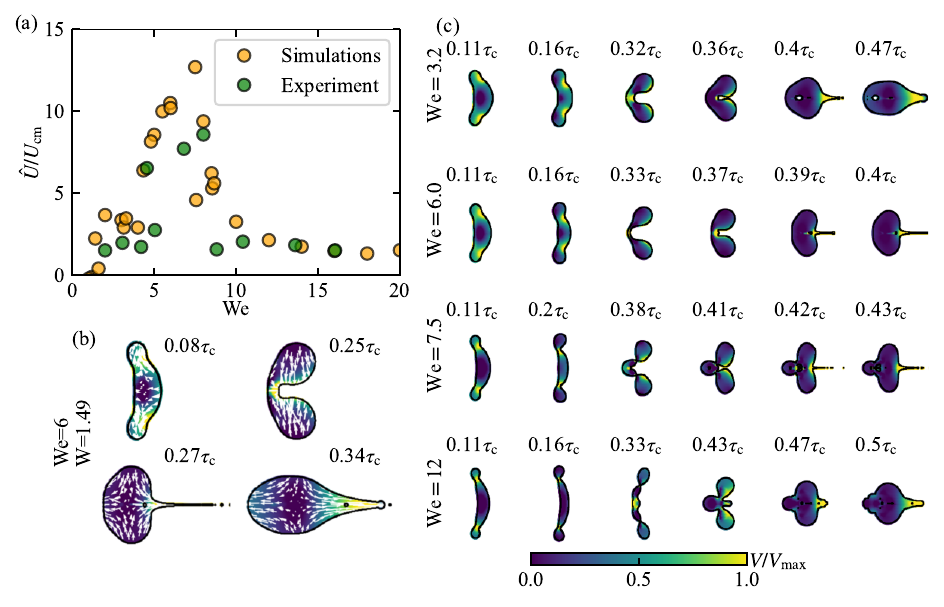}
    \caption{Simulation results for singular jetting. (a) Variation of the dimensionless jet velocity $\hat{U}/U_\mathrm{cm}$ over $\We$ for experiments (green squares), and simulations (orange circles). 
    (b) Four frames of the singular jetting before and after cavity collapse. The white arrows depict the local flow velocity, showing the bidirectional flow upon retraction. (c) Numerical frames of four examples of droplet jetting and the underlying cavity dynamics at different $\We$ and scaled values of $\W$ from the relation $\W\sim\We^{0.1}$. From left to right, several fractions of capillary times $\tauc$ are presented. From top to bottom: At $\We=3.2$, low jetting velocity caused by smooth cavity collapse and subsequent bubble entrapment. At $\We=6.0$, the combination of an increased radial flow and symmetric cavity collapse leads to the singular jet. The strongest jet is observed at $\We=7.5$ caused by asymmetric cavity breakdown, leading to the subsequent bubble entrapment. At $\We=12$ sheet-like expansion is observed, with $D_\mathrm{max}\sim2D_0$, and a slower jet is produced after retraction.}
    \label{figure:velocity_vs_diameter}
\end{figure}

We start the discussion of the main results by illustrating three different cases of droplet deformation with varying laser pulse energy in Figs.\,\ref{fig:presentation}(c-e) leading to different values of $\We$. After laser impact, the droplet acquires two orthogonal characteristic velocities, namely, center-of-mass propulsion $U_\mathrm{cm}$, and radial expansion rate $\dot{R}_0$. For the case of a nanosecond laser pulse and a beam size similar to the diameter of the droplet, the corresponding Weber numbers are similar $\We\sim\Wed$ (see discussions on the kinetic energy partitioning between the radial and center-of-mass degrees of freedom) under similar experimental conditions\,\cite{reijers2017, hernandez2022early, francca2025laser}). From here on, we focus on $\We$ to characterize the droplet dynamics upon laser impact. 
In Figs.\,\ref{fig:presentation}(c-e), we increase $\We$ from top to bottom and display five different frames in fractions of capillary time $\tau_c$. In each case, we compare the experimental data (black shapes) with simulations (colored shapes) to gain further insight into the jetting mechanism. The color map depicts the normalized magnitude of the velocity.
Note that the experimental frames are side views, whereas frames from the simulations present cross-sections of the droplets. 
Although hollow features cannot be captured in the experiment, the two representations still lead to nearly identical outer contours.
To match simulations with experiments, we input $\We$ with the corresponding value $\W$ using the scaling relation established in Ref.\,\cite{kharbedia2025_LWO}: $\W=\left(60\We^{0.1}-3\right)/48$. Since this scaling relation involves a degree of uncertainty, we allow some fine-tuning of the values of $\W$ (while remaining within $10\%$ of the predicted value).
In Figs.\,\ref{fig:presentation}(c-e), we display the time evolution of the jet after laser impact.
After retraction of the initial radial deformation, the radial flow leads to horizontal extension. When the inertia overcomes the capillary tension, i.e. $\Weu>1$, the droplet breaks and sheds fragments [Fig.\,\ref{fig:presentation}(c)]. In Appendix\,\S\,\ref{app:breakup_regimes} we provide a broader picture of the droplet breakup, as a function of the horizontal expansion rate, including the transition from dripping to jetting\,\cite{clanet_transition_1999}.  In this low Weber number regime, the retracting droplet has a pronounced curvature. Due to this curvature, and the predominant flow of the liquid along the laser propagation axis (left to right), after retraction, a small cavity is trapped within the droplet [see frame at $0.5\tauc$ in Fig.\,\ref{fig:presentation}(d) just prior to such trapping]. The subsequent collapse of this cavity leads to singular jetting. A suitable combination of the curvature and radial flow leads to a particularly thin and fast jet, as shown at $\We=8.0$ [Fig.\,\ref{fig:presentation}(d), at $0.64\tauc$]. The maximum jetting velocity observed in the experiments is around $15$\,m/s for $D_0 = 70\,\mu\mathrm{m}$, which is several times larger than the center-of-mass propulsion speed, reaching up to $\hat{U}\sim 10\,U_\mathrm{cm}$. This observation aligns very well with previously reported studies on the enhanced jetting mechanism after droplet impact on non-wetting solid surfaces, in a similar Weber number range \,\cite{bartolo_singular_2006, chen_submillimeter-sized_2017, guo_oblique_2020, zhang2022impact}. When further increasing the pulse energy, the radial expansion starts to dominate [cf. Fig.\,\ref{fig:presentation}(e) at $\We=10$].
The capillary tension flattens the recoiling droplet, avoiding effective cavity entrapment. The result is a slower and thicker jet that, as we will show, is not born from singularity.

In Fig.\,\ref{figure:velocity_vs_diameter}(a), we analyze in detail the jet velocity as a function of $\We$. To compare experiments with numerical simulations, we plot the dimensionless jet velocity $\hat{U}/U_\mathrm{cm}$ with $\We$. The jet velocity is obtained by performing a linear fit to the location of the tip of the jet over $0.1\tauc$ after it extended beyond the original right-hand edge of the spherical droplet (similar to Ref.\,\cite{chen_submillimeter-sized_2017, guo_oblique_2020}), where it systematically becomes traceable in the experiment. We observe that the jet velocity increases monotonously with $\We$, up to a maximum around $\We\sim8$, after which it starts to decrease to finally reach a low plateau. As argued above, the enhanced jetting regime is an outcome of the interplay between radial flow and sufficient forward curvature of the retracting droplet. The former continuously increases with $\We$, but the latter decreases with increasing $\We$. Consequently, there exists an optimum range, $\sim6<\We<8$, where both effects maximally boost the fast thin jet. The presence of the cavity can be confirmed solely from the simulations. From the experiments, we can only infer its entrapment by comparing the late-time evolution of the jet, and the corresponding velocity at similar values of $\We$. From Fig.\,\ref{figure:velocity_vs_diameter}(a), we confirm a similar trend for experiments and simulations.
To carry on a systematic analysis of the simulations, we filter spurious fragments (a manifestation of the finite mesh resolution) that are below two pixels in size. 
We find good agreement between simulation and experiment, supporting the original hypothesis that the fast jetting is a result of the infinite curvature at cavity collapse: we demonstrate singular jetting in a free-falling droplet.

A key distinction in the case of free-flying droplets is the absence of a contact line with a solid substrate. Given the radial distribution of the near-instantaneous plasma-induced pressure on the droplet, it naturally acquires a curvature that promotes cavity formation during retraction. The morphology of the retracting droplet depends on the pressure distribution width, which is again directly linked to the propulsion Weber number $\We$ \cite{kharbedia2025_LWO}.  
A closer look at the cavity across a wide range of $\We$ values reveals a highly non-trivial behavior. In Fig.\,\ref{figure:velocity_vs_diameter}(c) we show a time series of numerical results for different $\We$ along the curve displayed in Fig.\,\ref{figure:velocity_vs_diameter}(a). At $\We=3.2$, the droplet barely expands radially and shows a remarkable axial flow ($U_\mathrm{ax}>\dot{R}_0$). The resulting morphology leads to the formation of a central cavity that smoothly collapses. Although the cavity is present, the insufficient radial flow leads to a slow jet. The condition $U_\mathrm{ax}>\dot{R}_0$ induces an asymmetric collapse, with the rightmost part of the droplet closing before the bottom of the cavity. This leads to the formation and trapping of a (long-lived) spherical bubble, similar to previously reported studies\,\cite{bartolo_singular_2006, chen_submillimeter-sized_2017, guo_oblique_2020}. At $\We=6.0$, the recoil flow is fast enough to promote a thinner and faster jet, as clearly appreciated at $0.39\tauc$. Given a stronger radial flow, the collapse of the cavity is more symmetrical. Further increasing the radial flow, at $\We=7.5$, we observe the fastest jet. Notably, the collapse symmetry breaks again (see $0.38\tauc-0.41\tauc$), leading to the entrapment of a bubble. 
Unlike the low-$\We$ regime, the radial inflow here is strong enough to pinch the cavity, generating a bi-directional jet (note the small jet entering the bubble at $0.43\tauc$). The strongest jets observed correspond to this mechanism.
This behavior is consistent with the classical Worthington jet reported after the impact of a heavy object onto a liquid pool\,\cite{gekle_gordillo_2010}. Interestingly, a similar symmetry breaking in cavity collapse, governing the transition between collapses with and without bubble entrapment, has been observed in droplet impact experiments on non-wetting surfaces\,\cite{bartolo_singular_2006} as a function of impact velocity. However, in contrast to our observations, Bartolo et al.\,\cite{bartolo_singular_2006} reported the strongest jet following a symmetric collapse of the air-filled cavity. 
When radial flow is further increased, at $\We=12$, prior to retraction, the droplet acquires sheet-like structure, with $D_\mathrm{max}\gtrsim2D_0$. Greater radial expansion suppresses the effect of the cavity, and leads to a slower and thicker jet. 

Despite the high values of the singular jetting velocity around $\We\sim8$ with $\hat{U}/U_\mathrm{cm}\sim9$, it is smaller than observed in droplet impact experiments on non-wetting surfaces\,\cite{bartolo_singular_2006, chen_submillimeter-sized_2017, guo_oblique_2020}, where values up to $\hat{U}/U_\mathrm{cm}\sim20$ were reported. A possible explanation is the absence of solid surface that guides the flow normal to it after retraction. In free-falling droplets, the retraction leads to a horizontal expansion in both directions, as shown in Fig.\,\ref{figure:velocity_vs_diameter}(b). Consequently, the resulting jet velocity decreases. In Appendix \S\,\ref{app:velocity_comparison} we provide a direct comparison with the reported data on jetting velocity after droplet impact from some representative studies. Furthermore, given the strong curvature of the cavity within the singular jet regime, a mesh-resolution dependence is expected. We provide additional details on different mesh sizes and its influence on jetting dynamics in Appendix\,\S\,\ref{app:mesh_size}.

In conclusion, we validate our hypothesis, of a fast jetting mechanism based on cavity collapse, by numerical simulations. We further emphasize the complex behavior of the cavity within the droplet. In view of this non-monotonic dependence of the cavity dynamics on $\We$ (hence, on $\W$), we analyze its behavior as function of the these two parameters separately in the following section.

\subsection{Phase diagram}\label{sec:phase_diagram}

\begin{figure}
    \centering
    \includegraphics[width=1\textwidth]{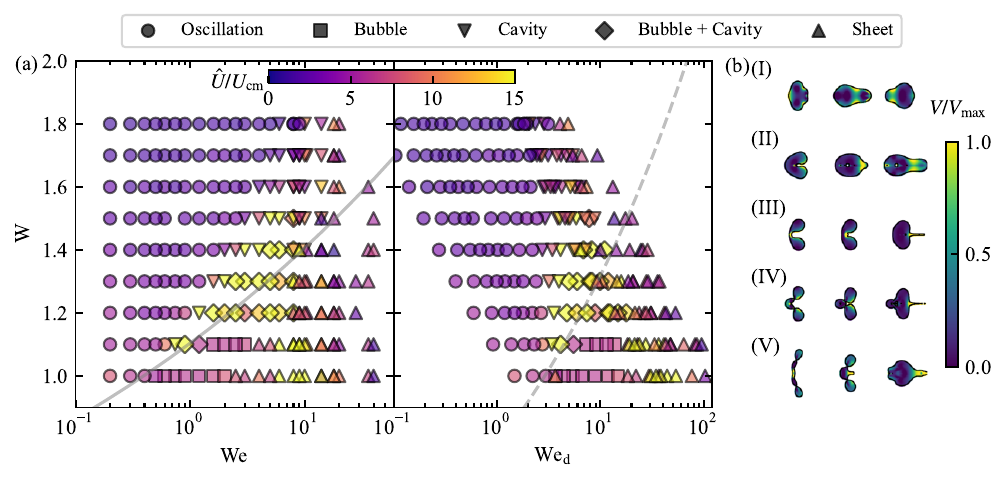}
    \caption{Phase diagram (simulations) of morphologies and jet velocity enhancement as a function of $\W$ and $\We$. (a) Simulated phase diagram relating $\W$ with $\We$ illustrating different jetting behaviors. Five relevant regimes have been identified: droplet oscillation (``Oscillation'', I), bubble entrapment (``Bubble'', II), symmetric cavity collapse (``Cavity'', III), asymmetric cavity collapse with subsequent bubble entrapment (``Bubble + Cavity'', IV), and sheet-like expansion (``Sheet'', V). Some representative frames for each case are depicted in (b). The gray line shows the empirical scaling that relates $\W$ and $\We$ under our experimental conditions (see main text). 
    }
    \label{fig:phase-diagram}
\end{figure}

Based on the dependence of the jetting dynamics on $\We$ and $\W$, we construct a phase diagram that spans a broad range of both parameters. Figure\,\ref{fig:phase-diagram}(a) presents a systematic exploration of the different morphologies observed in simulations, including (I) droplet oscillation, (II) bubble entrapment, (III) symmetric cavity collapse, (IV) asymmetric cavity collapse leading to bubble entrapment, and (V) sheet-like expansion at higher $\We$. Although bubble formation necessarily implies the prior presence of a cavity, we distinguish between the phases (II) and (IV), as the latter produces the fastest and most singular jets. 
The morphological classification is complemented by the corresponding jet velocities, represented as a color map. Representative frames for each phase are shown in Fig.\,\ref{fig:phase-diagram}(b). The highest velocities occur within a narrow range of $\We$ and $\W$ values (yellow region), highlighting the critical interplay between droplet curvature and radial flow. We complement this diagram with the empirical scaling law that relates $\We$ and $\W$, as discussed above. This curve corresponds to the experimental conditions, which are omitted from the diagram since the cavity shape cannot be identified experimentally. However, we have demonstrated good agreement between simulations and experiments on the jetting velocity as a function of $\We$ above in \S\,\ref{sec:cavity}. 
The strong sensitivity to the radial flow motivates translating this diagram into the framework of the deformation Weber number, $\Wed$, which directly quantifies the rate of radial deformation of the droplet, directly following Ref.\,\cite{kharbedia2025_LWO} based on an approach similar to that of Ref.\,\cite{klein_drop_2020}. Furthermore, $\Wed$ is linked to $\We$ through kinetic energy partition in laser-droplet interaction (see Refs.\,\cite{klein_drop_2020, hernandez2022early}). Therefore, it is possible to correlate the three parameters, namely, $\Wed$, $\We$, and $\W$ within the same scaling law (see Ref.\,\cite{kharbedia2025_LWO}): $\Wed\sim15\W^{-5}\We$, capturing the whole phase behavior in terms of a single parameter: the radial flow (i.e., $\Wed$). 
In the rightmost panel of Fig.\,\ref{fig:phase-diagram}(b), the singular jetting regime is well characterized by the threshold values of $\Wed$, with the leftmost boundary approaching a vertical line. Nevertheless, an additional sub-regime of singular jetting appears at the lowest $\W$ values, where the droplet behaves as a thin sheet ($20<\Wed<50$). Numerical simulations reveal highly curved sheets that fold during retraction, producing intense jets. This behavior is discussed in more detail in Appendix\,\S\,\ref{app:curved_sheets}.
Interestingly, at high $\Wed$ values the jetting behavior exhibits a non-monotonic dependence on $\W$. When the radial expansion becomes significant, surface capillary waves (CW) are generated near the bounding rim and converge radially as the sheet retracts. The interaction between these CW and the droplet curvature gives rise to complex cavity formation and collapse, often resulting in multiple jetting events. This phenomenon is analyzed independently in the following section to underscore the intricate nature of laser-induced jetting dynamics in free-falling droplets.

\subsection{Stepwise jetting: The role of capillary waves}\label{sec:stepwise}

\begin{figure}
    \centering
    \includegraphics[width=1\textwidth]{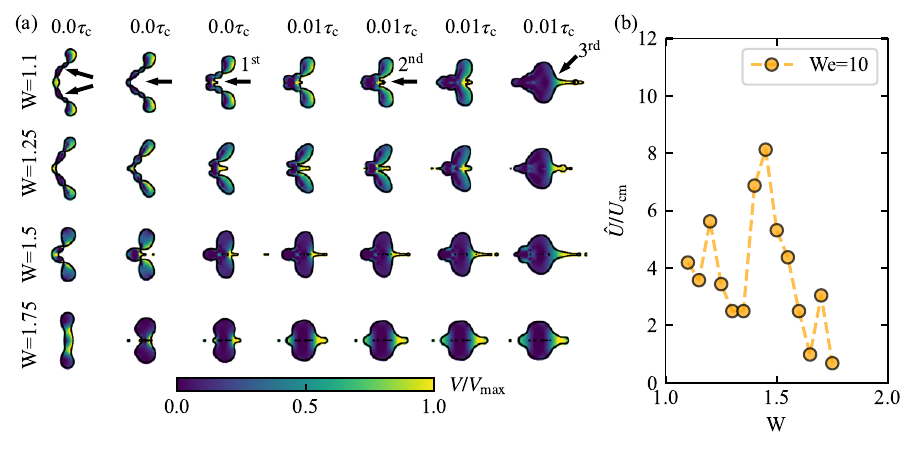}
    \caption{Simulation study of the cavity collapse and surface waves beyond the singular jetting regime, $\We>10$. (a) Time evolution of the cavity collapse (from left to right) at $\We=10$ and for different $\W$ values (top to bottom). The presence of CW is indicated with arrows at $0.26\tauc$. The converging CW induces stepwise cavity collapse during the droplet retraction, leading to three subsequent jets ($1^\mathrm{st},2^\mathrm{nd}$ and $3^\mathrm{rd}$). The number of observed waves decreases for higher $\W$ values, as shown for $\W=1.25,\,1.5\,\mathrm{and}\,1.75$. (b) Variation of the dimensionless velocity $\hat{U}/U_\mathrm{cm}$ with $\W$, corresponding to data in (a).}
    \label{fig:stepwise_CW}
\end{figure}

When the radial expansion becomes significant ($D_\mathrm{max} \gtrsim 2D_0$), the retraction phase exhibits a set of radially converging surface capillary waves (CW). These waves are particularly noticeable at low $\W$ values, where they induce stronger radial flows\,\cite{francca2025laser}, when the sheet becomes significantly thin and large. The presence of these waves constitutes an additional parameter influencing the jetting dynamics.
As the CW converge, they generate a sequence of small cavities. Combined with the inward radial flow, this process produces a step-wise jetting behavior. Figure\,\ref{fig:stepwise_CW}(a) shows time sequences of the droplet retraction for different $\W$ values at $\We=10$. In the first frame of the upper row, the CW are indicated by arrows. As the droplet retracts and the waves converge, a small cavity forms and collapses, producing the first jet at time $t=0.31\tauc$. The emerging jet and recoiling droplet then create a second cavity, which eventually collapses to form the next jet at $0.32\tauc$. Finally, as the droplet continues to retract, the remaining cavity closes and produces the main jet, as shown at $0.35\tauc$.
It is important to note that the velocity analyzed in this study always corresponds to the main jet, being the one that grows sufficiently to surpass the droplet's initial perimeter (the $3^\mathrm{rd}$ jet in Fig.\,\ref{fig:stepwise_CW}). Moreover, Fig.\,\ref{fig:stepwise_CW} shows that at higher $\W$ values, CW are not apparent. The occurrence of these waves results in a non-monotonic variation of the jetting velocity with $\W$, as illustrated in Fig.\,\ref{fig:stepwise_CW}(b) for $\We = 10$. Furthermore, each $\We-\W$ combination gives rise to distinct jetting dynamics. Under particular conditions of retraction rate and the phase of converging CW, the bubble entrapped after asymmetric cavity collapse can lead to a backward jetting, as shown for $\W=1.25$ at $0.33\tauc$ (see also Appendix\,\ref{app:curved_sheets}). From the last frame for each $\W$ value, we can appreciate visually the oscillatory variation of the jetting velocity with increasing values of $\W$. 

\section{Conclusions}

We report on singular jetting in free-falling tin droplets following the impact of a nanosecond laser pulse. The rapidly expanding plasma generated at impact exerts an instantaneous pressure impulse that can be characterized by the Weber number $\We$ and the dimensionless pressure width $\W$, which describes the spatial distribution of the applied pressure on the droplet's surface. After impact, the droplet undergoes rapid radial expansion and subsequently retracts, ultimately producing a jet.
Using numerical simulations performed with the Basilisk code, in tandem with the experiments, we show that a delicate interplay between the radial flow and the curvature of the retracting droplet leads to the formation of a cavity. For relatively low values of the Weber number ($\We<10$), the droplets exhibit a pronounced forward flow, which induces cavity formation after retraction. The collapse of this cavity dramatically amplifies the jet velocity, resulting in a jet that appears only within a narrow window of $\We=6-8$, reminiscent of the singular jets observed during droplet impact on solid substrates. The morphology of the resulting cavity is highly sensitive to both $\We$ and $\W$, and this sensitivity ultimately governs the jetting dynamics.
Unlike canonical singular-jet experiments involving droplet impact on substrates, here jet formation can be deliberately controlled in the simulations by selecting appropriate values of $\We$ and $\W$. This tunability enables the construction of a phase diagram in the $(\We,\W)$ space that organizes the various cavity morphologies. We identify distinct phases that lead to singular jetting, including both (forward-backward) symmetric and non-symmetric cavity collapses, the latter preceded by bubble entrapment. The phase diagram reveals a well-defined region of singular jetting that occurs only within a narrow range of $\We$ and $\W$, offering a straightforward mechanism for generating singular jets without the need for solid substrates or liquid pools.
Furthermore, at particularly low values of $\W$, corresponding to tightly focused pressure fields, and in the presence of sufficient radial flow, we identify an additional phase characterized by a highly curved sheet with radially converging capillary waves. The dynamics of these waves during cavity collapse profoundly influence the resulting jet velocity. Taken together, these results reveal a non-trivial yet fully controllable mechanism of jet formation, accessible through appropriate tuning of the laser parameters.

There are several immediate directions for future work. In this study, we define a singular jet primarily based on its thin morphology and unusually high velocity; however, a more careful analysis of the early-time dynamics of jet formation may reveal deeper mathematical and physical signatures of singular behavior, such as inertial self-similarity and collapse-driven focusing~\cite{gordillo2023theory_jets}. The simulations presented here follow experiments performed under vacuum conditions; the role of ambient gas pressure on jet dynamics therefore remains an open and practically important question, with relevance well beyond nanolithography. Moreover, a theoretical prediction for the critical Weber number We$\approx$8, which emerges here as it does in droplet-impact studies, is still lacking. From the perspective of nanolithography applications, additional open questions arise at the very earliest stages of plasma formation and droplet propulsion, where the values of W and We are initially set. Future studies that resolve these early-time processes may enable more reliable predictions of later-time phenomena, including the jet formation analyzed in this work.

\section{Acknowledgments}

The Advanced Center for Nanolithography (ARCNL) is a public-private partnership between the University of Amsterdam (UvA), Vrije Universiteit Amsterdam (VU), Rijksuniversiteit Groningen (UG), the Dutch Research Council (NWO), and the semiconductor equipment manufacturer ASML. This project was partly financed by 'Toeslag voor Topconsortia voor Kennis en Innovatie (TKI)' from the Dutch Ministry of Economic Affairs and Climate Policy, the project 'Plasma driven by a variable-wavelength laser for next-generation EUV sources for nanolithography' (with project no. 19458) of the Open Technology Programme (which is financed by NWO), and the European Research Council (CoG Project no. 101086839).

\newpage

\appendix\label{appendix}
\section{Jetting velocity: Comparison with literature}\label{app:velocity_comparison}

\begin{figure}
    \centering
    \includegraphics[width=0.5\textwidth]{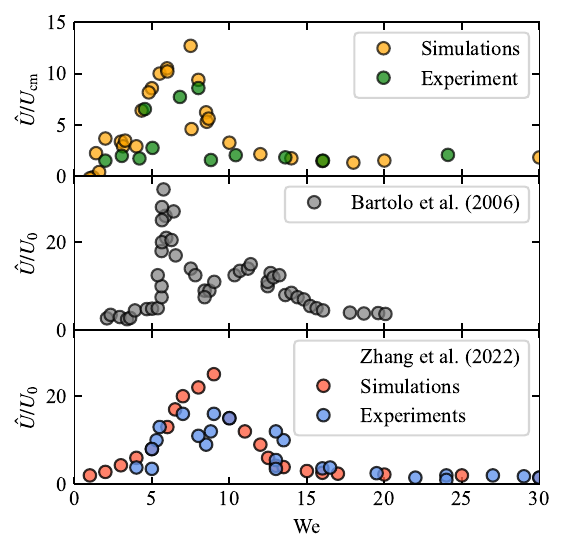}
    \caption{Comparison of the dimensionless jetting velocity $\hat{U}/U_\mathrm{cm}$ over $\We$ from this study (simulations and experiments) and reported data in the literature for water droplets with $D_0=2\,\mathrm{mm}$ after orthogonal impact on hydrophobic surfaces\cite{bartolo_singular_2006} (black circles) and performed on non-wetting substrates\,\cite{zhang2022impact} (red and blue circles). In case of droplet impact experiments, the jetting velocity is rescaled with the impact velocity of the droplet, $\hat{U}/U_0$. 
    Note the similarity in the range of $\We$ for the singular jetting that lies around $\We\sim6-8$ for all cases. The secondary peak at $\We\sim12$ observed in Ref.\,\cite{bartolo_singular_2006} is attributed to upward migration and collapse of an entrapped bubble (see the main text).}
    \label{fig:velocity_comparison}
\end{figure}

In this section we compare the singular jetting velocities from free-flying droplets from the current work with those that resulted from water droplet impact on a non-wetting surface. 
In our study, the maximum velocity of the singular jet turns out to be lower (for both experiment and simulation) than in case of droplet impact on non-wetting orthogonal surfaces (also both in experiments\,\cite{bartolo_singular_2006} and simulations\,\cite{zhang2022impact}). As demonstrated previously in section \S\,\ref{sec:phase_diagram}, the curvature of the retracting droplet sensitively influences the jet velocity. Although in both cases a non-trivial cavity collapse causes the enhanced jetting dynamics, in free-flying droplets, given the absence of the contact line, the momentum upon droplet retraction is channeled into the horizontal axis in both directions equivalently, conserving momentum [see Fig.\,\ref{figure:velocity_vs_diameter}(b)]. Consequently, the jet velocity would be expected to be lower when compared with droplet impact on a solid substrate. 
The singular jetting regime ranges similar Weber number values, within $6<\We<8$. Despite the complexity of the enhanced jetting, in both free-flying and impacting droplets, cavity collapse is the main driving force for the singular jet. In contrast, capillary waves (CW) often play the dominant role in cavity formation in droplet impact experiments\,\cite{bartolo_singular_2006, thoroddsen2018singular_jet}. 
Nevertheless, CW-mediated strong jetting can also be caused by a significantly curved sheet, as demonstrated in Fig.\,\ref{fig:phase-diagram}(a) and discussed in section\,\S\,\ref{fig:stepwise_CW}.  
Finally, in case of droplet impact, experiments show an additional peak at higher Weber values, $10<\We<15$. This secondary peak is attributed to the upward migration of the entrapped bubble after the asymmetric cavity collapse and its subsequent burst on the surface\,\cite{bartolo_singular_2006,chen_submillimeter-sized_2017,guo_oblique_2020}. We highlight further differences between free-flying and impacting droplets by noticing that the bubble, whenever entrapped after the recoil, migrates backwards [toward the laser-facing side; see Fig.\,\ref{app:curved_sheets}]. Therefore, the entrapped bubble burst is directed toward the laser. 
These differences between laser-droplet and droplet-impact cases notwithstanding, we find a remarkable similarity in terms of both the maximally achievable jet velocity, and the range of low We numbers at which it occurs.

\section{Singular jetting in highly curved sheets}\label{app:curved_sheets}

\begin{figure}
    \centering
    \includegraphics[width=0.6\linewidth]{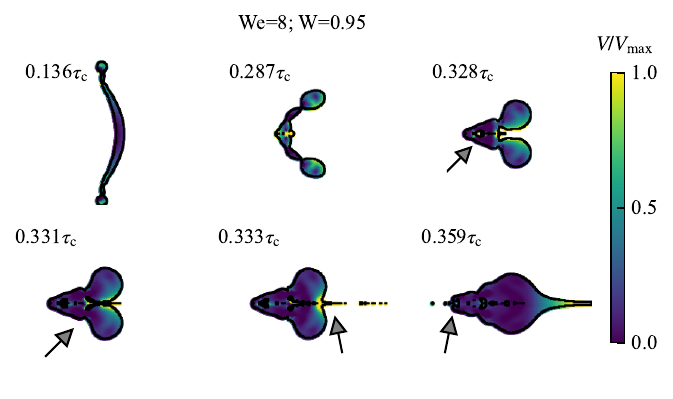}
    \caption{Strongly curved sheet and the resulting singular jetting. Different frames that depict the complexity of jetting dynamics at low pressure width values, $\W=0.95$ and $\We=8$. The remarked curvature is clearly appreciated at $0.287\tauc$. The confluence of curvature and CW leads to multiple cavity formations and subsequent collapses, as indicated with gray arrows at $0.328\tauc$ and $0.331\tauc$. The singular jet is observed at $0.333\tauc$. The migration of the bubble resulted from the first cavity collapse (at $0.328\tauc$) toward the front side of the droplet leads to surface pinch at $0.359\tauc$.}
    \label{fig:curved_sheets}
\end{figure}

The phase diagram portrayed in Fig.\,\ref{fig:phase-diagram} displays a significant increase in jetting velocity for tightly focused pressure fields, i.e., at low $\W$ values.
Contrary to what we would expect in sheet-like retracting droplets, such tightly focused pressure fields, $\W<1$, lead to pronounced curvature of thin sheets bounded by thick rims, as observed at $0.136\tauc$ in Fig.\,\ref{fig:curved_sheets}. Subsequently, the sheet folds and creates a massive cavity, as can be clearly seen at $0.287\tauc$ in the same figure. The presence of CW leads to successive cavity collapses, as indicated with gray arrows at times $0.328\tauc$ and $0.331\tauc$. These collapses result in the entrapment of two bubbles of different sizes. Once the two rims collide, the singular jet is observed at $0.333\tauc$. Finally, the migration of the first entrapped bubble towards the front of the droplet produces a small pinch on the surface, resulting in a jet, as observed at time $0.359\tauc$. Note the small time steps depicted in this regime. Interestingly, this effect smooths as we increase $\We$, where the radial flow becomes strong enough to flatten the sheet before retraction (see Fig.\,\ref{fig:phase-diagram}). 

\section{Mesh size influence on jetting dynamics}\label{app:mesh_size}

\begin{figure}
    \centering
    \includegraphics[width=1\linewidth]{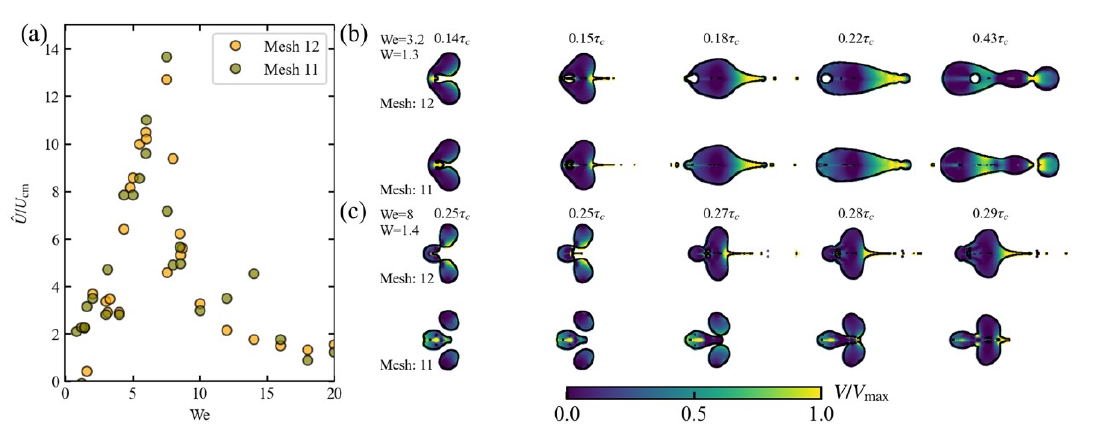}
    \caption{(a) Dimensionless jetting velocity $\hat{U}/U_\mathrm{cm}$ over propulsion Weber number $\We$ for two different mesh resolutions. (b) Comparison of the droplet morphology for two mesh resolutions at $\We=3.2$ and (c) at $\We=8$ (with the corresponding $\W$ values shown) for various times after laser impact. }
    \label{fig:mesh_size}
\end{figure}

In this section, we briefly assess the influence of mesh resolution on the jetting dynamics. Specifically, we compare the refinement level of ``mesh 12'' (a 4096x4096 grid resolution) used to obtain the results in the main text, with the lower-resolution ``mesh 11'' (2048x2048). As shown in Fig.\,\ref{fig:mesh_size}(a), the two meshes produce broadly similar jet behaviors, with a nearly identical peak velocity confined to the same range $\We=6-8$. For both grid resolutions, fragments smaller than two pixels were discarded. The remaining minor discrepancies between the two curves are attributed to subtle differences in cavity morphology. Moreover, filtering out only the elements smaller than two pixels may be insufficient to eliminate spurious fragments, stemming from numerical artifacts, during velocity estimation. Consequently, the velocity curves in Fig.\,\ref{fig:mesh_size} exhibit a non-monotonic variation of $\hat{U}/U_0$ with respect to $\We$. Reducing the mesh resolution further prevents reliable reproduction of singular jetting. 
Next, we compare two jets produced at $\We=3.2$ and $\We=8$, simulated with the different mesh sizes, as shown in Figs.\,\ref{fig:mesh_size}(b) and (c), respectively. At the lower Weber number ($\We = 3.2$), the droplet morphologies are qualitatively similar, especially at early times ($0.14\tauc$). However, finite-size effects become significant once the cavity forms and collapses, processes that involve extremely large curvatures, leading to subtle resolution-dependent differences that ultimately produce distinct jetting dynamics. These effects are clearly visible at $0.43\tauc$: the lower-resolution mesh exhibits fragment splitting, whereas in the higher-resolution case the fragment remains attached to the droplet. 
The differences become more pronounced near the singular jetting regime at $\We=8$, as illustrated in Fig.\,\ref{fig:mesh_size}(c). In the high-resolution simulation, the retracting droplet forms a cavity that drives the singular jet, while the low-resolution simulation instead exhibits splitting along the thin parts connecting the two rims (see the frames at $0.25\tauc$). This premature disconnection dramatically alters the subsequent jetting dynamics, as observed at later times.

\section{The role of viscosity}\label{app:viscosity}

\begin{figure}[htbp]
    \centering
    \includegraphics[width=0.5\textwidth]{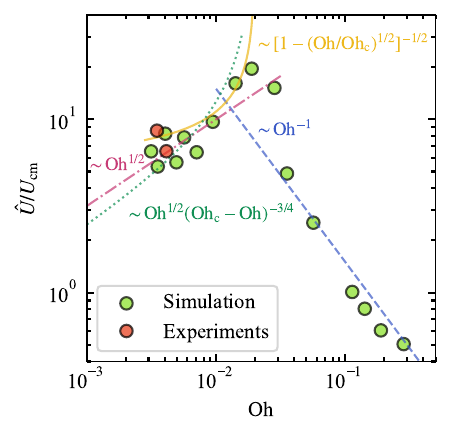}
    \caption{Comparison of the jetting velocity $\hat{U}/U_\mathrm{cm}$ over a wide range of Oh for $\We=8$ with various scaling laws available in the literature: blue dashed line from Deike et al.\,\cite{deike2018dynamics} and the others from Gordillo et al.\,\cite{gordillo2019capillary}.}
    \label{fig:viscosity}
\end{figure}

Here, we analyze the dependence of the normalized jetting velocity, $\hat{U}/U_\mathrm{cm}$, on the liquid viscosity by changing the Ohnesorge number, defined as $\mathrm{Oh}\equiv\sqrt{\mathrm{We}}/\mathrm{Re}=\mu/\sqrt{\rho \sigma D}$. In the numerical simulations, $\mathrm{We}$ is held constant while $\mathrm{Re}$ is varied, resulting in different values of $\mathrm{Oh}$. We note that in the experiments in the main text both We and Re are set to match the individual experimental conditions. Previous studies on Worthington-type jets generated by surface bubble bursting in Newtonian liquids\,\cite{deike2018dynamics,gordillo2019capillary,Sanjay2021} have reported a non-monotonic dependence of the jet velocity on $\mathrm{Oh}$, with a maximum at a critical value $\mathrm{Oh}_\mathrm{c}\sim0.02$, followed by a sharp decrease at higher $\mathrm{Oh}$ due to enhanced viscous dissipation. We observe a qualitatively similar behavior for laser-induced singular jets by analyzing the jetting velocity at $\mathrm{We}=8$ and varying $\mathrm{Re}$, hence $\mathrm{Oh}$, as shown in Fig.\,\ref{fig:viscosity}.
For $\mathrm{Oh}<\mathrm{Oh}_\mathrm{c}\approx 0.02$, the jetting velocity increases and reaches a maximum. In bubble bursting, the regime $< \mathrm{Oh_\mathrm{c}}$ has been attributed to inertia of capillary waves produced from the contracting rim. The viscosity dampens the longest wavelength $\lambda$ following scaling $\lambda\sim\mathrm{Oh}^{1/2}$. As $\mathrm{Oh}$ increases, the dominant wavelength of the capillary waves becomes larger, allowing inertial effects to overcome viscous dissipation and leading to an increase in jet velocity\,\cite{gordillo2019capillary}. Therefore, the convergence of the fastest waves result into increasing jet velocity, following  $\hat{U}\sim \mathrm{Oh}^{1/2}$ (red-dotted line in Fig.\,\ref{fig:viscosity}). Interestingly, Gordillo et al.\,\cite{gordillo2019capillary} observed that the shape of the cavity barely changes in this regime. Nevertheless, for $\mathrm{Oh}\sim\mathrm{Oh_c}$, a small bubble is entrapped within a conical cavity, and the inertia of the contracting walls leads to the scaling dependence $\hat{U}\sim[1-(\mathrm{Oh}/\mathrm{Oh_c}^{1/2})]^{-1/2}$ (orange curve in Fig.\,\ref{fig:viscosity}).
For $\mathrm{Oh}>\mathrm{Oh}_\mathrm{c}$, viscous dissipation dominates wave propagation across all wavelengths, resulting in a pronounced decrease of the jetting velocity, following $\hat{U}\sim\mathrm{Oh}^{-1}$ \cite{deike2018dynamics,gordillo2019capillary}. Although the role of capillary waves in laser-induced jets in free-falling droplets is not evident within the range $\mathrm{We}=6-8$ explored in this study, we observe a sharp reduction in jetting velocity at higher viscosities that is reminiscent of the bubble bursting case.

\section{Droplet breakup regimes}\label{app:breakup_regimes}

\begin{figure}[htbp]
    \centering
    \includegraphics[width=1\textwidth]{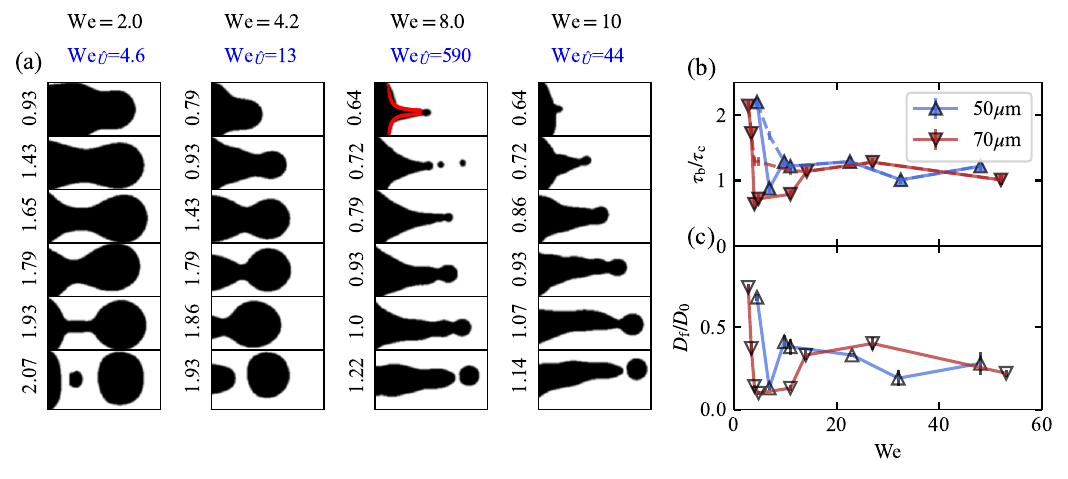}
    \caption{Study of jet breakup time and fragment size distribution. (a) A sequence of jet necking regimes over time (top to bottom) at different $\We$ (left to right). Several breakup regimes are shown. Droplet splitting at $\We=2.0$, dripping at $\We=4.2$, and jetting at $\We=8.0$ and 10. Frames are sequenced from the moment when the necking is observed up to the breakup. Cavity enhanced jetting is observed at $\We=8.0$, with the corresponding jetting Weber number $\Weu=590$. Note the sharp increase in $\Weu$ for $\We=8.0$. Also, two breakup processes are depicted in this case: The one corresponding to cavity collapse at $t/\tauc=0.72$ and late-time jetting at $t/\tauc=1.22$ (see the main text for discussion). The self-similar profile given by Eq.(\ref{eq:self-similar}) is shown for $\We=8.0$ at $0.64\tauc$. (b) Variation of the dimensionless breakup time $\tau_\mathrm{b}/\tauc$ as a function of $\We$ number for two different droplet diameters: $D_0=50$ and $70\mu\mathrm{m}$. The solid lines include fragmentation from cavity-enhanced jetting. The dashed lines depict the breakup instances excluding those instances. (c) Variation of the dimensionless fragment diameter $D_\mathrm{f}/D_0$ over $\We$. The studied fragment is the first one that is ejected upon jet breakup. Note the drastic decrease in breakup time (b) and fragment diameter (c) due to the cavity-enhanced jetting around $\We=8$.}
    \label{fig:breakup-regimes}
\end{figure}

Finally, we discuss a rather more general picture of the droplet deformation, where the jetting is a particular case of the droplet breakup. 
Experimentally, we observe that the breakup occurs only for $\Weu>1$. Depending on the horizontal flow, we expect to have different breakup processes depending on the control parameters $\We$ and $\W$. In Fig.\,\ref{fig:breakup-regimes}(a), we present several breakup regimes over time (top to bottom), and we focus on the rightmost part of the droplet, where a tip is visible. The increase in tip size suggests a continuous mass influx that facilitates the necking and subsequent detachment of a fragment. At $\We=2.0$, the low inertia of the tip ($\Weu=4.6$) leads to a slow $\sim\tauc$ necking. The amount of mass accumulated within the shed fragment is similar to the remaining droplet, $D_\mathrm{f}\sim D_0$. Consequently, the necking of the bridge in between is almost symmetrical, leading to the characteristic satellite droplet at $t=2.07\tauc$, similar to dripping liquid threads\,\cite{clanet_transition_1999,eggers_physics_2008}. 
Further increases in horizontal flow augment the inertia of the tip (higher $\Weu$), which tends to detach from the droplet earlier. At $\Weu=13$, a smaller fragment $D_\mathrm{f}< D_0$ is released at $t=1.93\tauc$. Right after, the remaining droplet retracts. Similarly to previous studies\,\cite{clanet_transition_1999}, the dripping regime precedes the jetting regime, and is a function of the flow rate. At $\We=8$ and $\We=10$, we observe the typical jetting regime, with the droplet continuously shedding much smaller fragments, $D_\mathrm{f}\ll D_0$. The cavity-enhanced jetting is clearly appreciated at $\We=8.0$ where we have a sharp increase in jetting velocity $\Weu=590$,. It decreases at $\We=10$ with $\Weu=44$. Interestingly, after cavity-induced jetting at $\We=8.0$ (see the $t=0.72\tauc$ example case in the figure), the droplet keeps feeding the tip, providing late-time breakup, as illustrated between $t=0.79-1.22\tauc$. The shedding dynamics becomes chaotic as we traverse from dripping to jetting regimes, in line with previous studies\,\cite{clanet_transition_1999}. For higher horizontal flows, at $\We=10$, the early-time cavity-driven jet is absent, and the fragmentation takes place much later. 

We therefore distinguish between two types of jetting, the one enhanced by the entrapped cavity collapse ($\We=8.0$ at $t=0.72\tauc$), and the usual flux-induced jetting ($\We=8.0$ at $t=1.22\tauc$, and $\We=10$). These two phenomena are not strictly delimited. 
Furthermore, the transition from dripping to jetting observed in our study qualitatively aligns with the dynamics of falling liquid threads, over similar ranges of the corresponding flow-based Weber number\,\cite{clanet_transition_1999,eggers_physics_2008}. The fundamental difference in our case is the presence of a cavity that introduces the anomaly in jetting velocity. This anomaly is appreciated whenever the breakup time, or the diameter of the shed fragments, is studied over $\We$, as illustrated in Figs.\,\ref{fig:breakup-regimes}(b) and (c), respectively. In Fig.\,\ref{fig:breakup-regimes}(b), we observe a sudden decrease in breakup time from low to medium $\We$ values, corresponding to the transition from dripping to jetting. To highlight the influence of the cavity-induced jetting, we compare the breakup instances with and without including the early-time jetting (solid and dashed lines, respectively). For high $\We$ values, the breakup reaches a plateau, and it is mainly driven by the Rayleigh-Plateau (RP) instability\,\cite{Rayleigh_instability_jet}. The corresponding breakup timescale of the fastest RP-growing mode can be quantified as $t_\mathrm{b}\sim2.91\sqrt{\rho D_\mathrm{f}/8\sigma}\sim\tauc$\,\cite{clanet_transition_1999, eggers_physics_2008,kharbedia2025rim}, in agreement with our experimental observations. Moreover, the size distribution of the first shed fragments follows a similar trend, as shown in Fig.\,\ref{fig:breakup-regimes}(c). 

For illustrative purposes, we further characterize the cavity-induced jet profile with the expected self-similar profile of slender jets, as derived for droplet and solid object impact on liquid pools\,\cite{van2021self-similar-jet}: 

\begin{equation}\label{eq:self-similar}
    R(z,t)=\frac{\sigma}{3\rho Kt}\left(\frac{z}{2t^{3/2}}+K\right)^{-1},
\end{equation}
where $K$ is a constant that is proportional to the ratio of surface tension and density; numerically,$K\approx1.5\pm0.5\mathrm{m}/\mathrm{s}^{3/2}$\,\cite{van2021self-similar-jet}. Given the non-trivial evolution of the cavity-induced jet in our study, Eq.\,(\ref{eq:self-similar}) can be used in a limited range of $\We$ and $t/\tauc$, where we clearly observe (in our simulations) a cavity entrapment. Moreover, its application requires a precise determination of the base of the jet, which is not feasible in our experimental data (see discussion in section \S\,\ref{sec:cavity}). Nevertheless, if we consider the approximate jet base to be the contour of the droplet at very early time of jetting $t<0.1\tauc$, Eq.\,(\ref{eq:self-similar}) reproduces well the jet profile, as illustrated in Fig.\,\ref{fig:breakup-regimes}(a) at $t=0.64\tauc$ for $\We=8.0$. At later times, after the horizontal mass flux induces the late-time jetting, the jet profile increasingly deviates from the predicted self-similar profile.

\newpage
\bibliographystyle{aipnum4-2}
\bibliography{bib_files/my_library}
\end{document}